\documentclass[reprint, twocolumn, superscriptaddress,prb]{revtex4-2}
\usepackage{bm, amsmath, amsfonts, amssymb, braket}
\usepackage{subfigure}
\usepackage{color}
\usepackage{graphicx}
\usepackage{slashed}

\usepackage{dcolumn} 

\usepackage{rotating} 

\usepackage{ulem}
\usepackage{xcolor}
\DeclareRobustCommand{\erase}{\bgroup\markoverwith{\textcolor{red}{\rule[.5ex]{2pt}{0.4pt}}}\ULon}

\begin{document}
\title{Localized-basis formulation of interacting Hamiltonians in flat topological bands: coherent states and coherent-like states for fractional physics}
\author{Nobuyuki Okuma}
\email{okuma@hosi.phys.s.u-tokyo.ac.jp}
\affiliation{%
  Graduate School of Engineering, Kyushu Institute of Technology, Kitakyushu 804-8550, Japan
}%

\date{\today}
\begin{abstract}
In topological bands, it is impossible to construct exponentially localized Wannier functions while preserving the symmetries. Instead, in quantum Hall systems, one can define an overcomplete basis of spatially localized coherent states.
In this work, we propose a unified framework for understanding the quantum Hall effect and Chern insulators from the perspective of localized bases, by extending the overcomplete basis of coherent states to Chern bands in terms of coherent-like states.
Specifically, by representing both coherent states and coherent-like states as wave packets defined on a band, the difference between them can be encoded solely in the functional form of the wave packet in momentum space. Furthermore, for filling factor $\nu=1/3$, we define a local repulsive interaction Hamiltonian based on these bases and discuss properties of its ground states. In particular, by relating this Hamiltonian to previously studied models, we show that in quantum Hall systems it possesses exactly zero-energy ground states with topological degeneracy, thereby confirming that it serves as a model for fractional quantum Hall systems. In addition, we numerically verify that the Hamiltonian possesses topological degeneracy for representative Chern insulator models. An advantage of this formulation is that it allows fractional quantum Hall systems and various fractional Chern insulator systems to be discussed within a unified framework using the same Hamiltonian form.
In addition, we discuss that coherent-like states can also be defined in $\mathbb{Z}_2$ topological insulators. Corresponding to the fermionic time-reversal symmetry of the system, Kramers-degenerate coherent-like states can be naturally defined. The localized basis constructed from coherent-like states is expected to be useful for describing strongly correlated topological phases in flat-band systems.

\end{abstract}
\maketitle
\section{Introduction}
In recent years, the study of topological phases has attracted considerable attention. Among these phenomena, the integer quantum Hall effect (IQHE) \cite{Klitzing-80} and the fractional quantum Hall effect (FQHE) \cite{FQHE-exp-82,Laughlin-Wavefunction-83,yoshioka-textbook-02,halperin-fractional-textbook-20,fradkin2013field} are regarded as foundational and pioneering examples.
From a theoretical standpoint, these effects are supported by the flatness of energy bands and the analytical properties of Landau levels, which has motivated extensive research from their discovery to the present day.
More recently, these concepts have been extended to topological insulators \cite{Kane-review,Zhang-review} and intrinsic topological order \cite{wen1990topological}.
While the former can be explained within a single-particle framework, the latter is fundamentally understood through many-body physics.

Among topological insulators, the Chern insulator~\cite{Haldane-88} represents the simplest two-dimensional example and can be regarded as topologically equivalent to the IQHE.
While the nontrivial topology in the IQHE arises from a strong magnetic field, that of the Chern insulator originates from the band structure itself and does not require a magnetic field.
Noting that the key distinction between the IQHE and FQHE lies in whether the filling factor $\nu$ is an integer or a fraction, one might naturally expect that a fractionally-filled Chern insulator would be topologically analogous to the FQHE.
A system in which an FQHE-like state emerges without an external magnetic field is referred to as a fractional Chern insulator (FCI) \cite{Regnault-Bernevig-11, Bergholtz-Liu-13, Parameswaran-13, Liu-Bergholtz-review-22}.
Although FCIs have not yet been fully established experimentally, recent studies have reported promising experimental signatures in two-dimensional materials \cite{li2021spontaneous,cai2023signatures,zeng2023thermodynamic,park2023observation,xu2023observation,lu2023fractional}.
However, this does not imply that all fractionally-filled Chern bands correspond to FCIs.
In general, Chern bands differ significantly from Landau levels, except in their topological characteristics.
For instance, Chern bands typically lack the flatness and analytical properties inherent to Landau levels.
Identifying materials that exhibit these properties at the non-interacting level constitutes a crucial first step in the search for FCIs, and considerable research has been devoted to this endeavor \cite{Regnault-Bernevig-11, Bergholtz-Liu-13, Parameswaran-13, Liu-Bergholtz-review-22, Parameswaran-Roy-Sondhi-12, Roy-geometry-14, Jackson-Moller-Roy-15, Claassen-Lee-Thomale-Qi-Devereaux-15, Lee-Claassen-Thomale-17, Mera-Ozawa-21-2, Varjas-Abouelkomsan-Kang-Bergholtz-22,ledwith2023vortexability,fujimoto2024higher}.
Other studies have focused on two-particle problems, in which interaction effects can be incorporated at a minimal level, and have suggested a connection between the dispersion \cite{Lauchli-Liu-Bergholtz-Moessner-13,Liu-Bergholtz-Kapit-13} and topology \cite{okuma2023relationship, OkumaMizoguchi2025} of bound states and FCIs.

In Chern insulators, Landau levels, and more general topological band structures, it is not possible to construct exponentially localized Wannier bases while preserving the symmetry \cite{thouless1984wannier,rashba1997orthogonal,thonhauser2006insulator,soluyanov2011wannier}. Therefore, real-space descriptions on periodic lattices are relatively rare. However, in quantum Hall systems, there exist some pioneering studies that use coherent states \cite{ishikawa1992field,ishikawa1999field,fremling2014coherent}.
The single-particle states in the lowest Landau level are defined as eigenstates of the harmonic oscillator under the symmetric gauge, where the ladder operators are given by projecting the complex coordinates $z,z^*$ onto the lowest Landau level. The eigenstates of this annihilation operator, i.e., the coherent states, are exponentially localized states in real space. While coherent states can be defined at an arbitrary coordinate point, they are not mutually orthogonal, and their set is overcomplete with respect to the lowest Landau level.
In our previous work, we defined coherent-like states that extend these coherent states to Chern insulators. In both cases, these states form a localized basis that can serve as a substitute for Wannier functions in topological systems, while also providing a framework for describing quantum Hall systems and Chern insulators on the same footing. In particular, in light of recent experimental progress on FCIs, it is meaningful to interpret FQHEs and FCIs in a unified manner.

In this study, we construct a minimal many-body Hamiltonian using special localized states defined in topological flat bands, namely, coherent states and coherent-like states.
We first reformulate the coherent-like states to describe their overcomplete basis structure. We then propose a Hamiltonian at filling $\nu=1/3$, in which states in these localized bases interact via local repulsive interactions. Although the Hamiltonian appears simple and can uniformly describe both FCIs and FQHEs, it possesses sufficient complexity to capture fractional physics owing to the non-orthogonality of the localized bases.
As evidence, by clarifying the connection to existing models, we show that this Hamiltonian possesses three degenerate exact zero-energy states in the lowest Landau level. Additionally, exact diagonalization is performed on representative Chern insulator models, confirming that the Hamiltonian exhibits threefold-degenerate gapped ground states, and we discuss conditions under which FCI states are stabilized in various Chern insulators.
The coherent-like states can also be extended to $\mathbb{Z}_2$ topological insulators, suggesting the potential for more general applications to strongly correlated topological flat-band systems.

This paper is organized as follows. In Sec. \ref{conventionsection}, we clarify subtle differences in conventions that often arise in band theory and introduce the conventions adopted in this paper together with the relevant notation. In Sec. \ref{coherent-like-section}, we define coherent-like states in Chern bands from the perspective of topological zero modes of projected vortex functions. In Sec. \ref{overcomplete-section}, we review overcomplete bases constructed from coherent states in a manner adapted to lattice systems and, inspired by this, define an overcomplete basis using coherent-like states. In Sec. \ref{hamiltonian-section}, we define interaction Hamiltonians formulated in terms of coherent states and coherent-like states, and show that they exhibit topological degeneracy characteristic of FQHEs and FCIs. In particular, for the FQHE, we demonstrate the existence of exact zero-energy ground states. In these Hamiltonians, the difference between quantum Hall systems and Chern insulators is encoded solely in the functional form of the wave packets that define the basis. In Sec. VI, we show that coherent-like states in $\mathbb{Z}_2$ topological insulators can be defined in a manner that reflects Kramers degeneracy. Finally, Sec. \ref{discussionsection} is devoted to a discussion of future directions.

\section{Convention and notation\label{conventionsection}}
In this section, we introduce the conventions and notations used in this paper.
As the kinetic part, we consider a fermionic quadratic lattice model given by:
\begin{align}
    \hat{H}=\sum_{\bm{R},\bm{R}'}\sum_{i,j}c^\dagger_{\bm{R},i}H_{(\bm{R},i),(\bm{R}',j)} c_{\bm{R}',j}, \label{tight}
\end{align}
where $\bm{R}=(X,Y)$ and $i,j=1,2,\cdots,n_{\rm orb}$ denote the unit cell vector and the intracell atomic orbitals, respectively.
Since our motivation is in solid-state physics, we impose discrete translation symmetry on Eq. (\ref{tight}).
As is well known, there are two common conventions for the Fourier transform:
\begin{align}
    &c^{\dagger}_{\bm{k},i}=\frac{1}{\sqrt{N_{\rm unit}}}\sum_{\bm{R}}e^{i\bm{k}\cdot\bm{R}}c^{\dagger}_{\bm{R},i},\label{indep}\\
    &\tilde{c}^{\dagger}_{\bm{k},i}=\frac{1}{\sqrt{N_{\rm unit}}}\sum_{\bm{R}}e^{i\bm{k}\cdot(\bm{R}+\bm{r}_i)}c^{\dagger}_{\bm{R},i},\label{position-dependent}
\end{align}
where $\bm{k}$ is the crystal momentum, $N_{\rm unit}$ is the number of unit cells, and $\bm{r}_i=(x_i,y_i)$ is the intracell sublattice position of orbital $i$. We adopt the convention (\ref{indep}). Under discrete translation symmetry, the Hamiltonian (\ref{tight}) can be rewritten as
\begin{align}
\hat{H}=\sum_{\bm{k}}\sum_{i,j}c^\dagger_{\bm{k},i}[H_{\bm{k}}]_{i,j}
    c_{\bm{k},j}.
\end{align}
The one-particle dispersion is obtained by diagonalizing the Bloch Hamiltonian matrix $H_{\bm{k}}$:
\begin{align}
    H_{\bm{k}}=\sum_{\alpha}\epsilon_{\bm{k},\alpha}\ket{u_{\bm{k},\alpha}}\bra{u_{\bm{k},\alpha}},
\end{align}
where $\alpha$ denotes the band index, and $\ket{u_{\bm{k},\alpha}}$ is the periodic parts of the Bloch eigenstates.
In the position basis $\{\ket{\bm{R},i}\}$, the Bloch eigenstates $\{\ket{\bm{k},\alpha}\}$ are represented as
\begin{align}
    \langle \bm{R},i\ket{\bm{k},\alpha}=u_{\bm{k},\alpha}(i)\frac{e^{i\bm{k}\cdot\bm{R}}}{\sqrt{N_{\rm unit}}},
\end{align}
where $u_{\bm{k},\alpha}(i):=\langle i\ket{u_{\bm{k},\alpha}}$. 
We assume that $\ket{u_{\bm{k},\alpha}}$ and $\ket{\bm{k},\alpha}$ are normalized.

In addition to the Fourier transform, there is also a convention issue regarding the sign of the Berry connection in band theory.
Here we define the Berry connection, Berry curvature, and Chern number as
\begin{align}
    a_I(\bm{k},\alpha)&=-i\bra{u_{\bm{k}},\alpha}\partial_{k_{I}}\ket{u_{\bm{k}},\alpha},\\
    \omega(\bm{k},\alpha)&=\frac{\partial a_y(\bm{k},\alpha)}{\partial k_x}-\frac{\partial a_x(\bm{k},\alpha)}{\partial k_y},\\
    C_{\alpha}&=\int_{\rm BZ} \frac{d^2k}{2\pi}\omega(\bm{k},\alpha),
\end{align}
where $I=x,y$, and BZ denotes the Brillouin zone.
Note that the Berry connection and curvature depend on the choice of the Fourier convention:
\begin{align}
    A_I(\bm{k},\alpha)&=-i\bra{U_{\bm{k}},\alpha}\partial_{k_{I}}\ket{U_{\bm{k}},\alpha}\notag\\
    &=a_I(\bm{k},\alpha)-\sum_{i}r^I_iu^*_{\bm{k}}(i)u_{\bm{k}}(i),\label{posi-berry}\\
    \Omega(\bm{k},\alpha)&=\frac{\partial A_y(\bm{k},\alpha)}{\partial k_x}-\frac{\partial A_x(\bm{k},\alpha)}{\partial k_y},
\end{align}
where $\ket{U_{\bm{k},\alpha}}$ is the periodic part of a Bloch eigenstate in the convention (\ref{position-dependent}).
Nevertheless, topological numbers such as the Chern number, which are globally defined, do not depend on the convention.

\section{Coherent-like states defined by vortex function\label{coherent-like-section}}
In this section, we first introduce the exact zero mode of the (Hermitian conjugate of) projected vortex function whose origin is topological. By using the exact zero mode, we define the coherent-like states on the lattice where the model is defined, which are analogous to the coherent states on the von Neumann lattice.
This section is a reorganization of a part of our previous work \cite{okuma2024constructing,okuma2025biorthogonal} with some additional elements.

\subsection{Lattice vortex function}
In the position basis $\{\ket{\bm{R},i}\}$, the lattice vortex function is represented as \cite{okuma2024constructing}
\begin{align}
    Z&=\alpha_1~ (X+\sum_i\tilde{x}_iP_i)+\alpha_2~ (Y+\sum_i\tilde{y}_iP_i)\notag\\
    &=\alpha_1~ X+\alpha_2~ Y +\sum_i\gamma_iP_i,\label{vortexfunction}
\end{align}
where $\alpha_1,\alpha_2,\gamma_i\in\mathbb{C}$ are parameters, and $P_i:=\ket{i}\bra{i}$ is the projection operator onto sublattice $i$. 
Physically, $\tilde{\bm{r}}_i=(\tilde{x}_i,\tilde{y}_i)$ represents a virtual sublattice position, which can differ from the real sublattice position.
The lattice vortex function is a direct analogy of the vortex function introduced for continuous models, which is a nonlinear extension of the complex coordinate $z=x+iy$ in the QHE context \cite{ledwith2023vortexability}.
In our lattice case, $Z$ is a diagonal matrix rather than a function.
If we set $(\alpha_1,\alpha_2)=(1,i)$ and $\tilde{\bm{r}}_i=\bm{r}_i$, Eq. (\ref{vortexfunction}) becomes a lattice analogue of the the complex coordinate $z=x+iy$.
The parameters of the lattice vortex function $\alpha_1,\alpha_2,\gamma_i\in\mathbb{C}$ should be determined based on some criteria, such as ideal conditions \cite{Roy-geometry-14,Jackson-Moller-Roy-15} and vortexability \cite{ledwith2023vortexability}.
If the determined $\tilde{\bm{r}}_i$ is different from the original position $\bm{r}_i$, the difference describes the non-uniformity in the spatial metric, which is discussed in terms of the $r$-ideal Chern band in continuous models \cite{estienne2023ideal}.
Alternatively, this can be regarded as the lattice counterpart of the nonlinearity inherent in the vortex function of the continuous model.

\subsection{Projected vortex function}
In the following, we focus on the physics projected onto a single Chern band $\alpha$ with Chern number $C>0$, whereas the extension to cases with multiple topological bands is straightforward \cite{okuma2024constructing}.
The projection operator onto the Chern band is given by
\begin{align}
    P=\sum_{\bm{k}}\ket{\bm{k}}\bra{\bm{k}}.
\end{align}
For simplicity, we will omit the band index $\alpha$ henceforth.
We first consider the infinite-volume limit with no boundary.
The wave functions in the Chern band are expressed in the wavepacket form:
\begin{align}
    \ket{\psi}\propto\int\frac{d^2k}{(2\pi)^2}a(\bm{k})\ket{\bm{k}}.
\end{align}
Equivalently, the coefficient $a(\bm{k})$ represents the wave function in the crystal-momentum basis.
In momentum space, the position operators (for the virtual configuration) are defined as the following differential operators:
\begin{align}
    P(X+\sum_i\tilde{x}_iP_i)P&\leftrightarrow \hat{x}_{\bm{k}}=i\partial_{k_x}-A_x(\bm{k}),\notag\\
    P(Y+\sum_i\tilde{y}_iP_i)P&\leftrightarrow \hat{y}_{\bm{k}}=i\partial_{k_y}-A_y(\bm{k}),
\end{align}
where $\bm{A}{\bm{k}}$ denotes the modified Berry connection depending on the virtual sublattice positions $\tilde{\bm{r}}_i$.
Due to the presence of the $P_i$ terms, $\bm{A}_{\bm{k}}$ corresponds to the convention (\ref{posi-berry}).
The modified Berry connection is calculated by replacing $\bm{r}_i$ in Eq. (\ref{posi-berry}) with $\tilde{\bm{r}}_i$.
The lattice vortex functions projected onto the Chern band are also represented by the following differential operators:
\begin{align}
    PZP&\leftrightarrow \hat{z}_{\bm{k}}=\alpha_{1} \hat{x}_{\bm{k}}+\alpha_2 \hat{y}_{\bm{k}},\\
    PZ^*P&\leftrightarrow \hat{z}^{\dagger}_{\bm{k}}=\alpha^*_{1} \hat{x}_{\bm{k}}+\alpha^*_2 \hat{y}_{\bm{k}},
\end{align}
which satisfy the commutation relation $[\hat{z}^{\dagger}_{\bm{k}},\hat{z}_{\bm{k}}]=2\Omega(\bm{k})$.
In the case of the lowest Landau level, the projected vortex functions (complex coordinates) in momentum space behave as ladder operators:
\begin{align}
    &\hat{z}^{(\mathrm{LLL})}_{\bm{k}}=\left[i\partial_{k_x}+\frac{k_y}{2\pi}-\partial_{k_y}\right],\\
&[\hat{z}^{(\mathrm{LLL})}_{\bm{k}}]^{\dagger}=\left[i\partial_{k_x}+\frac{k_y}{2\pi}+\partial_{k_y}\right],\\
&\left[\sqrt{\pi}[\hat{z}^{(\mathrm{LLL})}_{\bm{k}}]^{\dagger},\sqrt{\pi}\hat{z}^{(\mathrm{LLL})}_{\bm{k}}\right]=1,\label{lllcomm}
\end{align}
where we have used the Berry connection $\bm{A}_{\bm{k}}=(-k_y/2\pi,0)$ corresponding to the constant Berry curvature in the $C=1$ case, $\Omega(\bm{k})=1/2\pi$.

\subsection{Exact zero mode of projected vortex function\label{zeromodesection}}
For the construction of the coherent-like states, the following exact zero mode(s) are essentially important.
As shown in our previous work \cite{okuma2024constructing}, $PZ^*P$ has at least $C$ exact zero modes in the projected space. 
Here, the word ``mode" means a right eigenstate.
Let $\nu_{Z^*}$ and $\nu_{Z}$ be the numbers of exact zero modes of $PZ^*P$ and $PZP$ in the projected space, respectively.
According to our previous work \cite{okuma2024constructing}, the following relation holds:
\begin{align}
    \nu_{Z^*}-\nu_{Z}=C.
\end{align}
In the proof, we have used the Atiyah–Singer index theorem for a Dirac operator on the momentum-space torus:
\begin{align}
i\slashed{D}(\bm{k})=
    \begin{pmatrix}
        0&\hat{z}_{\bm{k}}\\
        \hat{z}^{\dagger}_{\bm{k}}&0
    \end{pmatrix}.\label{dirac}
\end{align}
Here, the gauge field appearing in the covariant derivative corresponds exactly to the Berry connection in momentum space.

For simplicity, we assume $C=1$ henceforth.
In this case, $PZ^*P$ possesses at least one exact zero mode.
Except in cases where $PZP$ has accidental zero modes, $PZ^*P$ possesses exactly one exact zero mode. In momentum-space representation, the coefficient of the zero-mode wavepacket, $a_0(\bm{k})$, satisfies
\begin{align}
\hat{z}^{\dagger}_{\bm{k}}a_0(\bm{k})=0.
\end{align}
Unlike the Wannier states in topological bands, the wavepacket corresponding to $a_0(\bm{k})$ is exponentially localized in real space.
In the case of the lowest Landau level, 
$[\hat{z}^{(\mathrm{LLL})}_{\bm{k}}]^{\dagger}$ acts as an annihilation operator, and the above zero mode is interpreted as the vacuum.
The explicit form of $a_0(\bm{k})$ is derived in Appendix \ref{xlqcoherent}:
\begin{align}
&[\hat{z}^{(\mathrm{LLL})}_{\bm{k}}]^{\dagger}a^{(\rm LLL)}_0(\bm{k})=0,\\
    &a^{(\rm LLL)}_0(\bm{k})=e^{-\frac{k_y^2}{4\pi}}\vartheta_3\left(\frac{-k_x+i k_y}{2\pi}|i\right),\label{a0lll}
\end{align}
where $\vartheta_3(z|\tau)$ denotes the elliptic theta function.

In an actual calculation of $a_0(\bm{k})$, it is preferable to consider the eigenvalue problem of the following Hermitian operators rather than the inherently unstable non-Hermitian operator:
\begin{align}
    PZPZ^*P~\mathrm{or}~\hat{z}_{\bm{k}}\hat{z}^{\dagger}_{\bm{k}}.
\end{align}
Evidently, $PZ^*P$ and $PZPZ^*P$ share the same zero mode.
The non-zero eigenmodes of $PZPZ^*P$ are also discussed in terms of the radially localized basis in our previous work \cite{okuma2024constructing}.
In the case $(\alpha,\beta)=(1,i)$, $\hat{z}_{\bm{k}}\hat{z}^{\dagger}_{\bm{k}}$ becomes
\begin{align}
    \left[(-i\partial_{k_x}+A_x)^2+(-i\partial_{k_y}+A_y)^2-\Omega(\bm{k})\right],\label{mom-space-landau}
\end{align}
where the Berry curvature is computed by replacing $\bm{r}_i$ in Eq. (\ref{posi-berry}) with $\tilde{\bm{r}}_i$.
This represents a Landau problem on a momentum-space torus with an additional potential term $\Omega(\bm{k})$. A similar momentum-space Hamiltonian operator was discussed in the context of momentum-space Landau levels \cite{Claassen-Lee-Thomale-Qi-Devereaux-15,Lee-Claassen-Thomale-17}, although it does not support an exact zero mode.

\subsection{Numerical calculation in finite periodic system\label{periodicnotation}}
In most applications of band theory, a finite system with periodic boundary conditions is considered an approximation of an infinite system without boundaries.
The vortex function, however, contains the position operators, which are incompatible with the periodic boundary condition.
Nevertheless, the following method works well.

Let us consider the operators in the Chern-band eigenbasis $\{\ket{\bm{k}}\}$, where the band index is omitted.
Then, the lattice vortex functions are represented as $N_{\rm unit}\times N_{\rm unit}$ matrices:
\begin{align}
    &PZP,PZ^*P\leftrightarrow\hat{Z},\hat{Z}^{\dagger},\notag\\
    &PZPZ^*P\leftrightarrow \hat{Z}\hat{Z}^{\dagger},\notag\\
    &[\hat{Z}]_{\bm{k},\bm{k}'}:=\bra{\bm{k}}Z\ket{\bm{k}'}.
\end{align}
Here, we consider a system in which the origin of the position coordinates is placed at the center, and periodic boundary conditions are imposed along all directions.
Instead of operators defined on the infinite system, one can use $\hat{Z}$ and $\hat{Z}^{\dagger}$.
Roughly speaking, the eigenstate of $PZPZ^*P$ corresponding to the $m$-th smallest eigenvalue is localized on an approximate circle centered at the origin, with radius proportional to $\sqrt{m}$.
Consequently, when considering eigenstates with small eigenvalues, boundary effects are negligible.
It should be noted that small-eigenvalue states can exist at the system boundaries, where the position operators are discontinuous.
Indeed, zero modes may appear at the corners of the system. Such artificial zero modes must be removed by an appropriate method. In this paper, we introduce a small potential term at the corner to $\hat{Z}\hat{Z}^{\dagger}$.

\begin{figure}[]
\begin{center}
 \includegraphics[width=8cm,angle=0,clip]{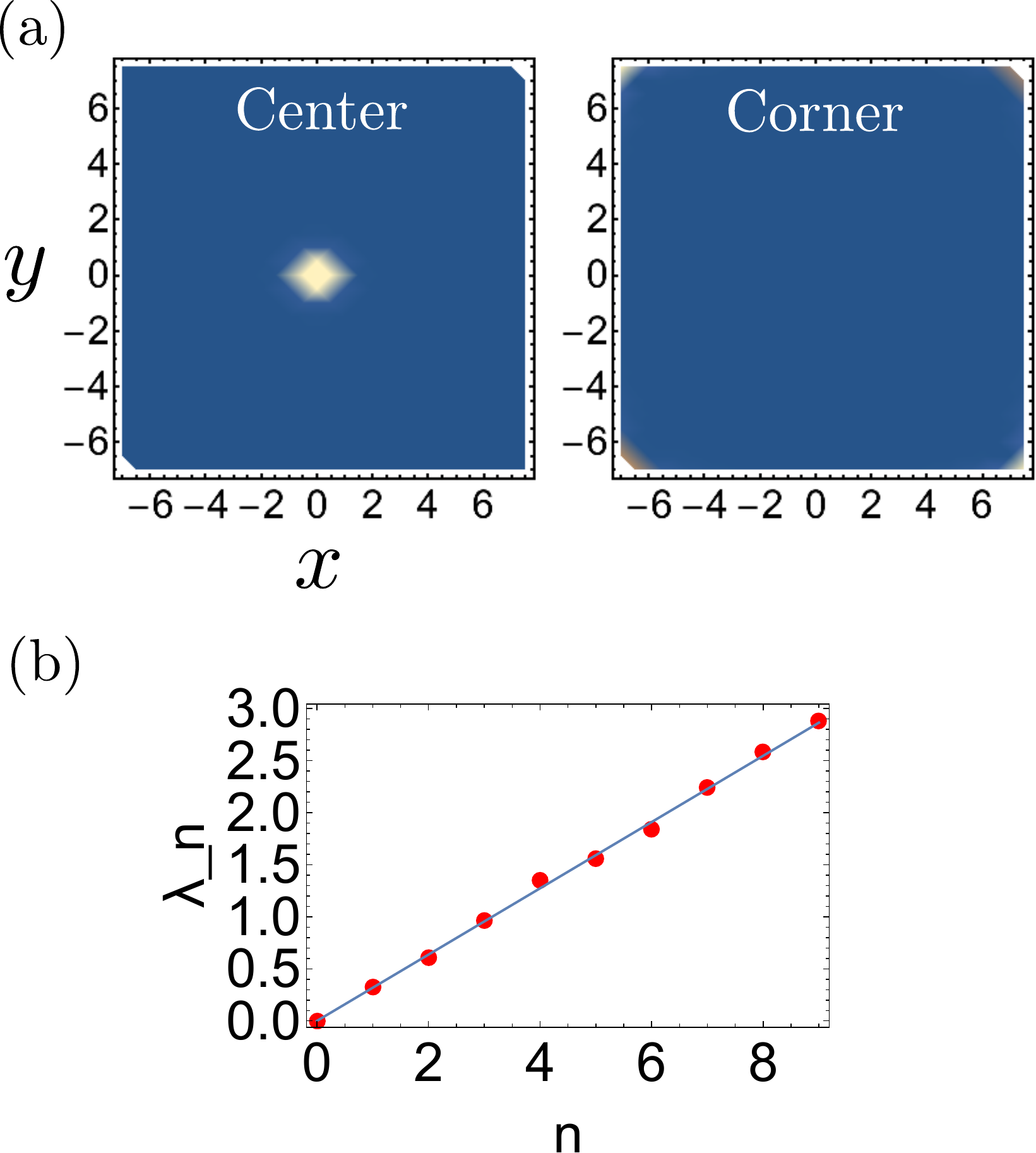}
 \caption{(a) Zero-modes of $PZPZ^*P$ in checkerboard lattice model. (b) The ten smallest eigenvalues of $\hat{Z}\hat{Z}^{\dagger}$, except for one quasi-zero mode. The blue line corresponds to $\lambda_n=n/\pi$.} 
 \label{fig1}
\end{center}
\end{figure}

As an example, we compute the real-space configurations of the zero modes of $PZPZ^*P$ for a model defined on the checkerboard lattice \cite{Neupert-Santos-Chamon-Mudry-11}.
This model is defined on a square lattice and includes two orbital degrees of freedom within the unit cell.
See Appendix \ref{models} for details.
We set $(\alpha_1,\alpha_2)=(1,i)$, $\tilde{\bm{r}}_1=(0.5,0)$, and $\tilde{\bm{r}}_2=(0,0.5)$, where the lattice constant is unity.
The system size $N_{\rm unit}$ is $15\times15$.
We find one (almost) exact zero mode ($\sim10^{-12}$) at the center and one quasi-zero mode ($\sim10^{-3}$) at the corner [Fig.\ref{fig1}(a)]. The latter zero mode is not essential for the present analysis. 

For the same setup, we also calculate the eigenvalues of $\hat{Z}\hat{Z}^{\dagger}$, except for one quasi-zero mode.
In the lowest Landau level, the eigenvalues of the counterpart of $PZPZ^*P$ are quantized as $\lambda_n=n/\pi$ since the system is described by a quantum harmonic oscillator. In terms of the eigenvalues, this model exhibits a similar behavior [Fig.\ref{fig1}(b)]. 
The eigenvalues of $PZPZ^*P$ and $PZZ^*P$ were also investigated in our previous work \cite{okuma2024constructing} and in Ref. \cite{Claassen-Lee-Thomale-Qi-Devereaux-15,Lee-Claassen-Thomale-17}, respectively.
In our previous work \cite{okuma2024constructing}, we calculated the eigenvalues of Eq.(\ref{mom-space-landau}) under a discretization, and the values are slightly different from the eigenvalues of $\hat{Z}\hat{Z}^{\dagger}$.
This difference is expected to vanish in the infinite-volume limit.

\subsection{Coherent-like state on von Neumann lattice}
We now reconsider the infinite system.
Actually, one can construct an infinite number of eigenstates of $PZ^*P$ because the following holds trivially:
\begin{align}
    \hat{z}^{\dagger}_{\bm{k}}[e^{-i\bm{k}\cdot\bm{R}}a_0(\bm{k})]&=[\hat{z}^{\dagger}_{\bm{k}}e^{-i\bm{k}\cdot\bm{R}}]a_0(\bm{k})\notag\\
    &=Z^*_{\bm{R}}[e^{-i\bm{k}\cdot\bm{R}}a_0(\bm{k})],\\
    PZ^*P\ket{\zeta_{\bm{R}}}&=Z^*_{\bm{R}}\ket{\zeta_{\bm{R}}},\\
    \ket{\zeta_{\bm{R}}}&\propto\int\frac{d^2k}{(2\pi)^2}e^{-i\bm{k}\cdot\bm{R}}a_0(\bm{k})\ket{\bm{k}},
\end{align}
where $Z_{\bm{R}}=\alpha_1X+\alpha_2Y$.
This is analogous to coherent states, which are eigenstates of the annihilation operator.
We refer to these states as coherent-like states.
In the lowest Landau level, where $[\hat{z}^{(\mathrm{LLL})}_{\bm{k}}]^{\dagger}$ acts as an annihilation operator, these states become coherent states in the strict sense.

For the true coherent states, complete subsets of the (overcomplete) set exist in an infinite system.
These subsets are characterized by a two-dimensional translation-invariant lattice, known as the von Neumann lattice \cite{von2018mathematical,perelomov2002completeness, bargmann1971completeness}.
Note that these subsets are complete, but not orthogonal.
The notion of the von Neumann lattice has been discussed in quantum Hall physics \cite{imai1990field,ishikawa1992field,ishikawa1999field}, where the lowest Landau level is described by creation and annihilation operators.
As an analogy, we regard the set $\{\ket{\zeta_{\bm{R}}}\}$ as a natural generalization of the coherent states on the von Neumann lattice.
In our case, the von Neumann lattice coincides with the periodic lattice on which the model is defined.

The coherent-like state is characterized by a coefficient $a_0(\bm{k})$, which has a zero point $\bm{k}_0$ in the Brillouin zone.
By adding a constant shift to the virtual sublattice positions $\{\tilde{\bm{r}}_i\}$, this zero point moves in the Brillouin zone because the constant shift means the flux insertion to the noncontractible loops of the momentum-space torus.
In the lowest Landau level, Eq.(\ref{a0lll}), the zero of $a^{(\rm LLL)}_0(\bm{k})$ is located at $\bm{k}=(\pi,\pi)$.
Note that the position of the zero point depends on the definition of $a^{(\rm LLL)}_0(\bm{k})$.
For example, Ref. \cite{ishikawa1999field} used $\vartheta_1$ whose zero point is located at $\bm{k}=\bm{0}$.

\begin{figure}[]
\begin{center}
 \includegraphics[width=8.2cm,angle=0,clip]{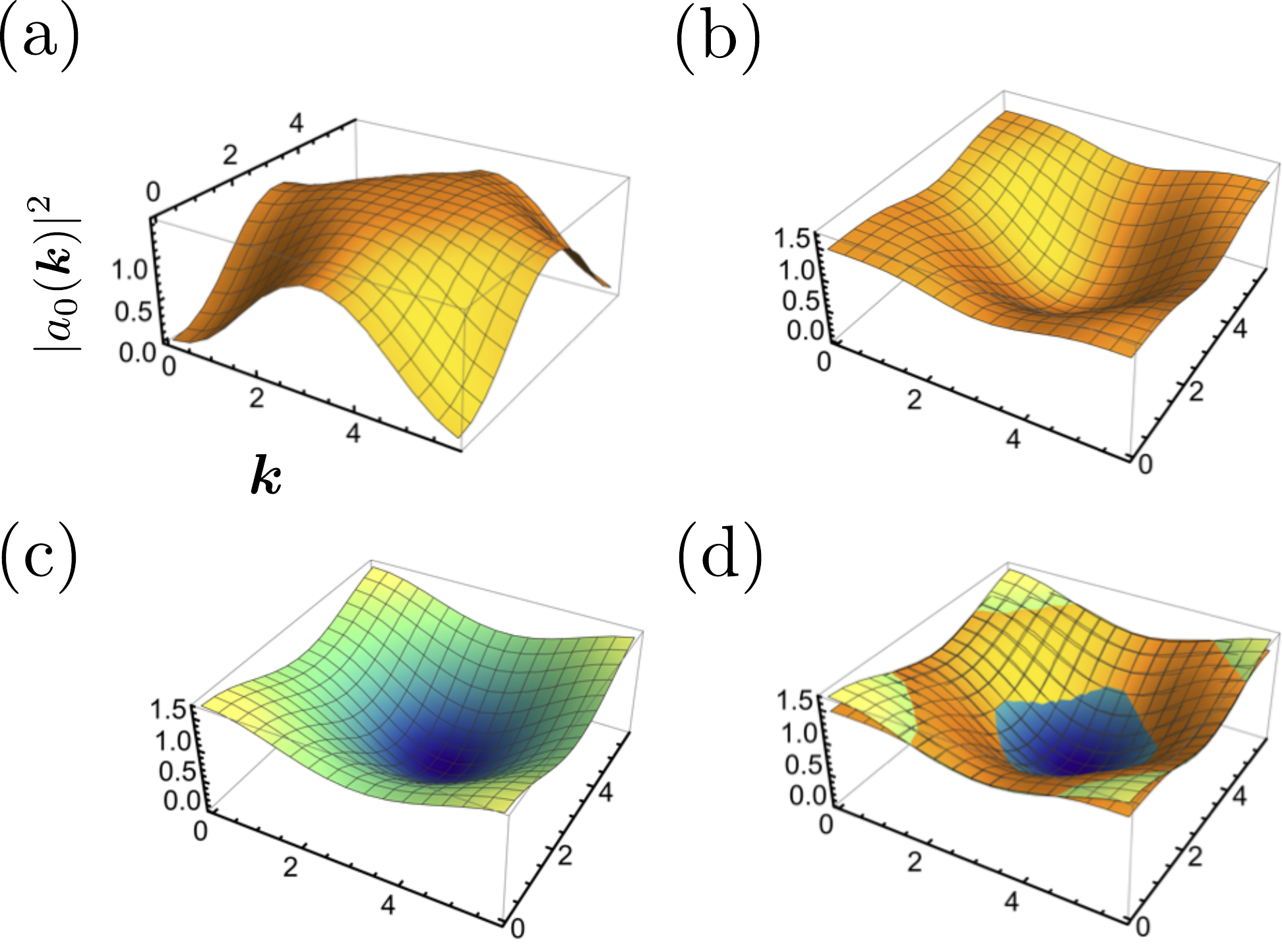}
 \caption{Distributions $|a_0(\bm{k})|^2$ for (a) $\tilde{\bm{r}}_1=(0.5,0),~\tilde{\bm{r}}_2=(0,0.5)$, (b) $\tilde{\bm{r}}_1=(0,-0.5),~\tilde{\bm{r}}_2=(-0.5,0)$. The counterpart in the lowest Landau level is shown in (c). The comparison between (b) and (c) is shown in (d).}
 \label{fig2}
\end{center}
\end{figure}

As an example in Chern insulators, we again examine the checkerboard-lattice model.
We calculate the distribution $|a_0(\bm{k})|^2$ for (a) $\tilde{\bm{r}}_1=(0.5,0),~\tilde{\bm{r}}_2=(0,0.5)$, (b) $\tilde{\bm{r}}_1=(0,-0.5),~\tilde{\bm{r}}_2=(-0.5,0)$.
Here, the constant shift is $-(0.5,0.5)$.
In both cases, one can find the zero point in the Brillouin zone [Fig.\ref{fig2}(a,b)].
While we have used $N_x=N_y=15$, the position of the zero point depends on the parity of $N_x$ and $N_y$.
We also compare $|a_0(\bm{k})|^2$ in case (b) with the lowest Landau-level counterpart [Fig.\ref{fig2}(c, d)].
In terms of the distribution $|a_0(\bm{k})|^2$, the checkerboard-lattice model is very similar to the lowest Landau level.

\subsection{States with higher angular momentum}
The operator $PZPZ^*P$ is an analogy of the angular-momentum operator in the lowest Landau level:
\begin{align}
    &P_{\rm LLL}zP_{\rm LLL}z^{*}P_{\rm LLL}\ket{\phi_{m}}=2m\ket{\phi_{m}},
\end{align}
where $m$ labels the angular momentum.
In the momentum-space language, the corresponding eigen equation in Chern insulators is given by
\begin{align}
    \hat{z}_{\bm{k}}\hat{z}^{\dagger}_{\bm{k}}a_{m}(\bm{k})=\lambda_m a_{m}(\bm{k}),\label{higher-angular}
\end{align}
where $m$ mimics the angular-momentum index, and $m=0$ corresponds to the zero mode, $\lambda_0=0$.
In the case of the lowest Landau level, these states are related to each other via $([\hat{z}^{(\mathrm{LLL})}_{\bm{k}}]^{\dagger},\hat{z}^{(\mathrm{LLL})}_{\bm{k}})$, which behave as the ladder operators [see Eq.(\ref{lllcomm})].
For example, the explicit form of the $m=1$ state in the lowest Landau level is given by
\begin{align}
    &a^{(\rm LLL)}_1(\bm{k})=\sqrt{\pi}\hat{z}_{\bm{k}}a^{(\rm LLL)}_0(\bm{k})\notag\\&=\frac{e^{-\frac{k^2_y}{4\pi}}}{\sqrt{\pi}}
    \left[k_y\vartheta_3 \left(\frac{-k_x+i k_y}{2\pi}|i\right)-i\vartheta_3'\left(\frac{-k_x+i k_y}{2\pi}|i\right)  \right],\label{firstlll}
\end{align}
where $\vartheta_3'(z|i):=d\vartheta_3(z|i)/dz$.
Needless to say, such a beautiful relation does not exist for a general Chern band.”

\section{Overcomplete basis formed by coherent-like states\label{overcomplete-section}}
In the above basis, the momentum corresponding to the zero of $a_{0}(\bm{k})$ cannot participate in the physics.
In this section, in order to treat all momenta on an equal footing, we introduce coherent-like states localized at different positions within the unit cell and construct a von Neumann lattice for each momentum. Using these states, we consider a decomposition of the identity in terms of an overcomplete basis.

\subsection{Overcomplete basis for lowest Landau level}
Before considering the Chern insulator, we reformulate the known results about the overcomplete basis in the lowest Landau level. 
As is well known, the coherent states are not orthogonal to each other. Clearly, the set $\{\ket{\alpha}|\alpha\in\mathbb{C}\}$ is overcomplete.
Nevertheless, the following decomposition of identity holds:
\begin{align}
    \hat{1}=\frac{1}{\pi}\int d^2\alpha\ket{\alpha}\bra{\alpha}.
\end{align}
In the following, we derive a similar decomposition for the lowest Landau level in terms of the coherent states:
\begin{align}
    P_{\rm LLL}z^*P_{\rm LLL}\ket{\zeta^{\rm(LLL)}_{\bm{R}+\bm{\delta}}}&=Z^*_{\bm{R}+\bm{\delta}}\ket{\zeta^{\rm(LLL)}_{\bm{R}+\bm{\delta}}}.
\end{align}
The unit cell is defined as a square region penetrated by one quantum of magnetic flux.
$\bm{\delta}$ is the shift vector inside the unit cell.
For each $\bm{\delta}$, the von Neumann lattice is defined.

As a concrete form of the coherent state, we use the definition given in Ref. \cite{qi2011generic}.
Although the original definition is on a cylinder, we can generalize it to torus boundary conditions.
In Appendix \ref{xlqmomentum}, we derive the momentum-space representation of this coherent state:
\begin{align}
&\ket{\zeta^{\rm(LLL)}_{\bm{R}+\bm{\delta}}}=\sum_{\bm{k}}\frac{e^{-i\bm{k}\cdot\bm{R}}}{\sqrt{N_{\rm unit}}}a^{\rm(LLL)}_{\bm{k}}(\bm{\delta},0)\ket{\bm{k}},\\
&a^{(\rm LLL)}_{\bm{k}}(\bm{\delta},0)=e^{-ik_y\delta_y}a^{(\rm LLL)}_0(\bm{k}+2\pi\bm{\delta}\times \hat{e}_3),\label{symmetry}\\
&[\hat{z}^{\rm(LLL)}_{\bm{k}}]^{\dagger}a^{\rm(LLL)}_{\bm{k}}(\bm{\delta},0)=(\delta_x+i\delta_y)^*a^{\rm(LLL)}_{\bm{k}}(\bm{\delta},0),
\end{align}
where $a^{(\rm LLL)}_0(\bm{k})$ is given in Eq. (\ref{a0lll}).
If we set $\bm{\delta}=(j/N_y,i/N_x)$ with $0\leq j<N_y$ and $0\leq i<N_x$, $\bm{k}+2\pi\bm{\delta}\times \hat{e}_3$ is also a crystal momentum.
Thus, $a^{(\rm LLL)}_{\bm{k}}(\bm{\delta},0)$ for each $\bm{\delta}$ has zero points at differnt crystal momentum, $(\pi,\pi)-2\pi\bm{\delta}\times \hat{e}_3$, under even $(N_x,N_y)$.
By using this set, we can define the decomposition of the projection operator:
\begin{align}
    P_{\rm LLL} =\sum_{\bm{R}}\sum_{\bm{\delta}}\frac{1}{N_yN_x}\ket{\zeta^{\rm(LLL)}_{\bm{R}+\bm{\delta}}}\bra{\zeta^{\rm(LLL)}_{\bm{R}+\bm{\delta}}}.\label{LLLdecomp}
\end{align}
This is consistent with the known fact (for example, see Ref. \cite{fremling2014coherent}).

\begin{figure}[]
\begin{center}
 \includegraphics[width=8.2cm,angle=0,clip]{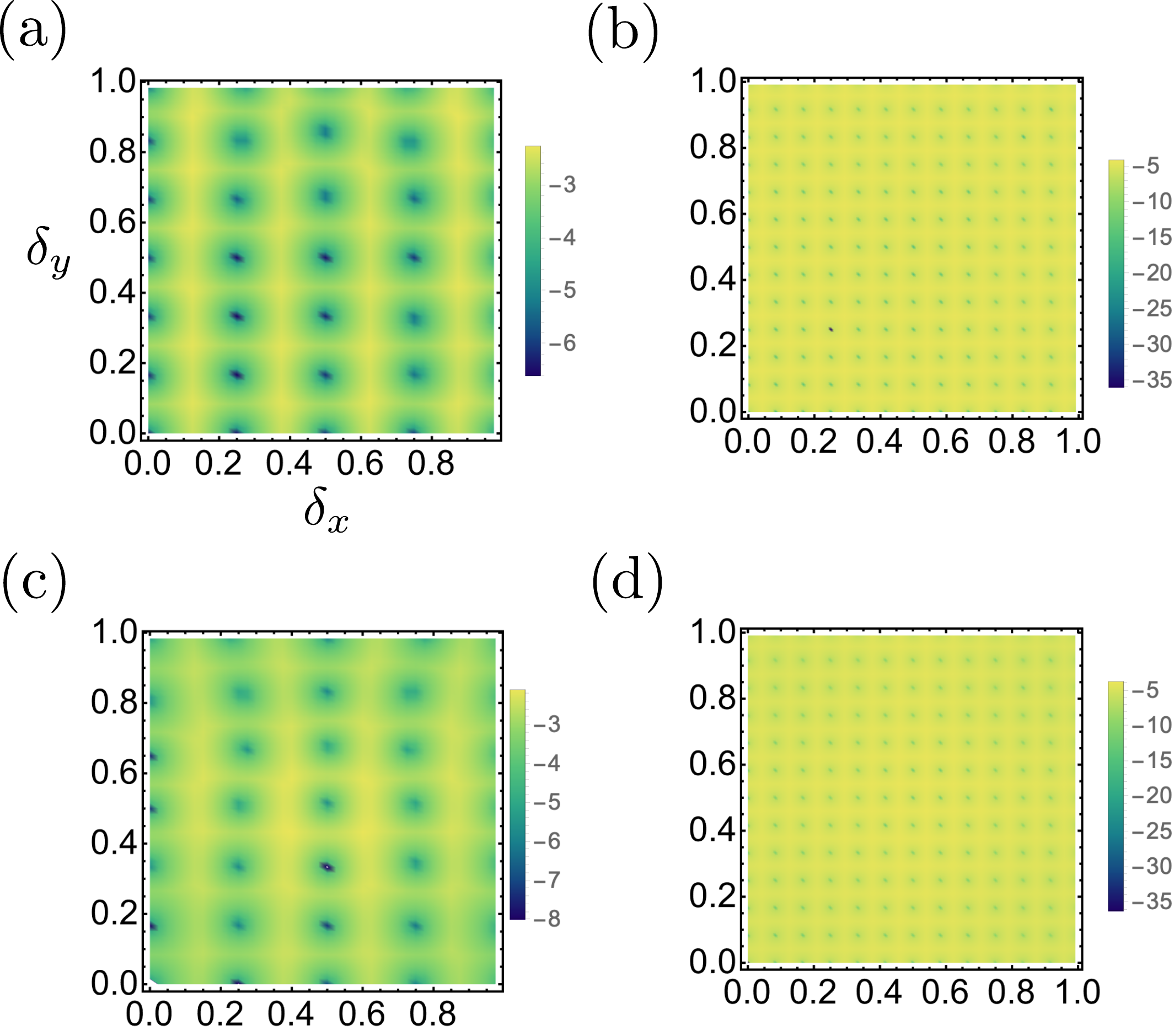}
 \caption{$\min_{\bm{k}}\log |a_{\bm{k}}(\bm{\delta},0)|$ as a function of $\bm{\delta}$ for the checkerboard-lattice model with (a) $N_x\times N_y=6\times4$ and $12\times12$, and for the QWZ model with (c)$N_x\times N_y=6\times4$ and $12\times12$. $\bm{\delta}=\bm{0}$ corresponds to  $\tilde{\bm{r}}_1=(0,-0.5),~\tilde{\bm{r}}_2=(-0.5,0)$ for the checkerboard-lattice model and $\tilde{\bm{r}}_1=(0,0),~\tilde{\bm{r}}_2=(0,0)$ for the QWZ model, respectively.}
 \label{fig3}
\end{center}
\end{figure}

\subsection{Overcomplete basis for Chern bands\label{overbasis}}
We here generalize the above discussion to Chern insulators.
In the following, we assume $(\alpha,\beta)=(1,i)$ and $C=+1$.
As we discussed, the constant shift in the lattice vortex function leads to a different zero mode, $a_0(\bm{k})$. Correspondingly, the von Neumann lattice is shifted in real space.
Instead of changing $Z$, we introduce a shift vector $\bm{\delta}\in[0,1)\times[0,1)$ to express the new coherent-like states:
\begin{align}
PZ^*P\ket{\zeta_{\bm{R}+\bm{\delta}}}&=Z^*_{\bm{R}+\bm{\delta}}\ket{\zeta_{\bm{R}+\bm{\delta}}},
\notag\\
\ket{\zeta_{\bm{R}+\bm{\delta}}}&=\sum_{\bm{k}}\frac{e^{-i\bm{k}\cdot\bm{R}}}{\sqrt{N_{\rm unit}}}a_{\bm{k}}(\bm{\delta},0)\ket{\bm{k}},\notag\\
\hat{z}^{\dagger}_{\bm{k}}a_{\bm{k}}(\bm{\delta},0)&=(\delta_x+i\delta_y)^*a_{\bm{k}}(\bm{\delta},0).
\end{align}
By rearranging the final expression, one finds that it is precisely the zero-mode equation for $Z$ shifted by a constant.
By introducing a $\bm{\delta}$-dependent factor $f(\bm{\delta})$, one can decompose the projection operator onto the Chern band: 
\begin{align}
    \hat{P}&=\sum_{\bm{R}}\sum_{\bm{\delta}}f(\bm{\delta})\ket{\zeta_{\bm{R}+\bm{\delta}}}\bra{\zeta_{\bm{R}+\bm{\delta}}},
\end{align}
where
\begin{align}
\sum_{\bm{k}}\ket{\bm{k}}\bra{\bm{k}}&=\sum_{\bm{k}}\sum_{\bm{\delta}}f(\bm{\delta})|a_{\bm{k}}(\bm{\delta},0)|^2\ket{\bm{k}}\bra{\bm{k}}\notag\\
\Leftrightarrow1&=\sum_{\bm{\delta}}f(\bm{\delta})|a_{\bm{k}}(\bm{\delta},0)|^2.
\end{align}
The fact that $f(\bm{\delta})$ is constant in the lowest Landau level originates from the symmetric form (\ref{symmetry}).

In the case of Chern insulators, the choice of $\bm{\delta}$ involves a subtle degree of arbitrariness.
In Fig.\ref{fig3}, we plot $\min_{\bm{k}}\log |a_{\bm{k}}(\bm{\delta},0)|$ as a function of $\bm{\delta}$.
As representative models, we investigated the checkerboard lattice model and the Qi–Wu–Zhang (QWZ) model \cite{Qi-Wu-Zhang-06}, with system sizes $N_x \times N_y = 6 \times 4$ and $12 \times 12$.
For each $\bm{\delta}$, $\arg \min_{\bm{k}}\log |a_{\bm{k}}(\bm{\delta},0)|$ gives the momentum numerically closest to the zero point.
The local minima in Fig. \ref{fig3} are located on the $N_y \times N_x$ lattice points, and for these values of $\bm{\delta}$, $a_{\bm{k}}(\bm{\delta},0)$ has a numerical (almost) zero point.
In the results with $N_x \times N_y = 6 \times 4$, the lattice is slightly distorted, whereas for $N_x \times N_y = 12 \times 12$ it is well aligned.
The slight distortion may be due to the finite-size effect.
In the following, we adopt $\bm{\delta}=(j/N_y,i/N_x)$ with $0\leq j<N_y$, $0\leq i<N_x$, as in the case of the lowest Landau level.
To avoid the effects of distortion, in calculations for small systems, we hereafter use the values of $a_{\bm{k}}(\bm{\delta},0)$ obtained from sufficiently large systems.

\subsection{Higher-angular-momentum states at arbitrary position}
For a shift vector $\bm{\delta}$, we can also define higher–angular-momentum states by replacing $(\hat{z}_{\bm{k}},\hat{z}^{\dagger}_{\bm{k}})$ with
$(\hat{z}_{\bm{k},\bm{\delta}},\hat{z}^{\dagger}_{\bm{k},\bm{\delta}})=[\hat{z}_{\bm{k}}-(\delta_x+i\delta_y),~\hat{z}^{\dagger}_{\bm{k}}-(\delta_x+i\delta_y)^*]$ in Eq. (\ref{higher-angular}).
Corresponding to the coherent-like states at $\bm{R}+\bm{\delta}$, we define local orbitals 
\begin{align}
&\ket{\eta^{(m)}_{\bm{R}+\bm{\delta}}}=\sum_{\bm{k}}\frac{e^{-i\bm{k}\cdot\bm{R}}}{\sqrt{N_{\rm unit}}}a_{\bm{k}}(\bm{\delta},m)\ket{\bm{k}},\notag\\
&\hat{z}_{\bm{k},\bm{\delta}}\hat{z}^{\dagger}_{\bm{k},\bm{\delta}}~a_{\bm{k}}(\bm{\delta},m)=\lambda_{m,\bm{\delta}}~ a_{\bm{k}}(\bm{\delta},m).
\end{align}
For $m=0$, it reduces to the coherent-like state, namely $\ket{\eta^{(0)}_{\bm{R}+\bm{\delta}}}=\ket{\zeta_{\bm{R}+\bm{\delta}}}$.
Using the $m$-th states, the projection operators can be decomposed as
\begin{align}
\hat{P}&=\sum_{\bm{R}}\sum_{\bm{\delta}}f^{(m)}(\bm{\delta})\ket{\eta^{(m)}_{\bm{R}+\bm{\delta}}}\bra{\eta^{(m)}_{\bm{R}+\bm{\delta}}}\notag\\
\Leftrightarrow1&=\sum_{\bm{\delta}}f^{(m)}(\bm{\delta})|a_{\bm{k}}(\bm{\delta},m)|^2.\label{mcomp}
\end{align}
Again, the explicit form for the lowest Landau level with $m=1$ is given by
\begin{align}
    a^{(\rm LLL)}_{\bm{k}}(\bm{\delta},1)=e^{-ik_y\delta_y}a^{(\rm LLL)}_1(\bm{k}+2\pi\bm{\delta}\times \hat{e}_3),
\end{align}
where $a_{1}^{(\rm LLL)}(\bm{k})$ is given in Eq. (\ref{firstlll}).

The corresponding creation operators are given by
\begin{align}
    c^{\dagger}_{\bm{R},\bm{\delta},m}=\sum_{\bm{k}}\frac{e^{-i\bm{k}\cdot\bm{R}}}{\sqrt{N_{\rm unit}}}a_{\bm{k}}(\bm{\delta},m)c^{\dagger}_{\bm{k}}.
\end{align}
Due to the Hermiticity of $\hat{z}_{\bm{k},\bm{\delta}}\hat{z}^{\dagger}_{\bm{k},\bm{\delta}}$, these local orbitals are orthogonal to each other.
Thus, the following anticommutation relation holds:
\begin{align}
    &\{c^{\dagger}_{\bm{R},\bm{\delta},m},c_{\bm{R},\bm{\delta},m'}\}=\delta_{m,m'},\label{commutation}\\
    &\{c^{\dagger}_{\bm{R},\bm{\delta},m},c^{\dagger}_{\bm{R}',\bm{\delta},m'}\}=\{c_{\bm{R},\bm{\delta},m},c_{\bm{R}',\bm{\delta},m'}\}=0.
\end{align}
Note that due to the non-orthogonality, $\{c^{\dagger}_{\bm{R},\bm{\delta},m},c_{\bm{R}',\bm{\delta}',m'}\}\neq0$ for $\bm{R}+\bm{\delta}\neq \bm{R}'+\bm{\delta}'$. 

\section{A simple interacting Hamiltonian on coherent-like states\label{hamiltonian-section}}
In this section, we define interaction Hamiltonians formulated in terms of coherent states and coherent-like states, and show that they exhibit topological degeneracy characteristic of FQHEs and FCIs. 
\subsection{Definition of the Hamiltonian}
We are now in a position to define the Hamiltonian analyzed in this paper.
Let us consider a fermion in the coherent-like states $\ket{\zeta_{\bm{R}+\bm{\delta}}}$.
This fermion experiences a repulsive interaction with fermions in other orbitals.
Assuming a short-range interaction, a minimal model describing such an interaction is given by
\begin{align}
    H=\sum_{\bm{R}}\sum_{\bm{\delta}}U(\bm{\delta})~n_{\bm{R},\bm{\delta},0}~n_{\bm{R},\bm{\delta},1},\label{coherent-hubbard}
\end{align}
where $n$ is the number operator for local orbitals, and $U(\bm{\delta})$ is a model parameter for describing the position-dependent interaction.
As mentioned above, the choice of $\bm{\delta}$ contains a subtle degree of arbitrariness.
We choose $\bm{\delta}=(j/N_y,i/N_x)$ with $0\leq j<N_y$ and $0\leq i<N_x$.

In momentum-space representation, the model can be written as
\begin{widetext}
\begin{align}
H=\frac{1}{N_{\rm unit}}\sum_{\bm{q},\bm{k},\bm{k}'}\sum_{\bm{\delta}}U(\bm{\delta})~
a_{\bm{k}}(\bm{\delta},0)a_{\bm{q}-\bm{k}}(\bm{\delta},1)a^*_{\bm{q}-\bm{k}'}(\bm{\delta},1)a^*_{\bm{k}'}(\bm{\delta},0)~c^{\dagger}_{\bm{k}}c^{\dagger}_{\bm{q}-\bm{k}}c_{\bm{q}-\bm{k}'}c_{\bm{k}'}.\label{scattering}
\end{align} 
\end{widetext}
Here, we use $\{c^{\dagger}_{\bm{R},\bm{\delta},0},c_{\bm{R},\bm{\delta},1}\}=0$ to obtain the normal-ordered form. 
This form is very similar to the conventional Hamiltonian with intra-unit-cell interactions projected onto the flat Chern band:
\begin{align}
    H=\frac{1}{2N_{\rm unit}}\sum_{i,j}&V_{i,j}~u^*_{\bm{k}}(i)u^{*}_{\bm{q}-\bm{k}}(j)u_{\bm{q}-\bm{k}'}(j)u_{\bm{k}'}(i)\notag\\
    &c^{\dagger}_{\bm{k}}c^{\dagger}_{\bm{q}-\bm{k}}c_{\bm{q}-\bm{k}'}c_{\bm{k}'}.
\end{align}
Namely, our model is realized by replacing the Bloch states $u_{\bm{k}}(i)$ of a conventional Hamiltonian with $a^*_{\bm{k}}(\bm{\delta},m=0,1)$.
More precisely, the factor $\sqrt{f^{(m)}(\bm{\delta})}$ from Eq. (\ref{mcomp}) should be included for normalization.
In this sense, our formalism provides a natural mapping from the original system to a system with $\mathcal{O}(N_{\rm unit})$ sublattice sites, which becomes nearly continuous for a large $N_{\rm unit}$.

\subsection{Exact zero-energy ground states for the lowest Landau level case}
Instead of the coherent-like states, we here consider the true coherent states $\ket{\zeta^{(\rm LLL)}_{\bm{R}+\bm{\delta}}}$ in the lowest Landau level.
For arbitrary $U(\bm{\delta})\geq0$ with $\bm{\delta}=(j/N_y,i/N_x)$, the model (\ref{coherent-hubbard}) possesses three zero-energy ground states that correspond to the FQHE.
In the following, we show this fact by relating the model (\ref{coherent-hubbard}) to a previously known model.
Rather than the model (\ref{coherent-hubbard}), we first consider the following Hamiltonian expressed as an integral:
\begin{align}
    H\propto\int dxdy ~n_{z,0}~n_{z,1}=\int dxdy c^{\dagger}_{z,1}c^{\dagger}_{z,0}c_{z,0}c_{z,1},\label{continuous}
\end{align}
where $z=(R_x+\delta_x)+i(R_y+\delta_y)$ denotes the complex coordinate.
In fact, this model is equivalent to the following model
\begin{align}
    H\propto\int dxdy ~\bm{b}^{\dagger}(z)\cdot\bm{b}(z),~\bm{b}(z)=c_{z,0}(-i\nabla)c_{z,0}.\label{localham}
\end{align}
Details can be found in Appendix \ref{proof1}.
This model on a cylinder geometry was studied in Ref. \cite{qi2011generic} and was shown to be equivalent to a second-quantized Hamiltonian of Haldane's pseudopotential for $\nu=1/3$, which has a unique gapped zero-energy ground state \cite{lee2004mott}.
Actually, the Hamiltonian (\ref{localham}) on a torus can be related to a similar model on a torus in Ref. \cite{ortiz2013repulsive}, which has three zero-energy gapped ground states corresponding to the topological degeneracy. See Appendix \ref{proof} for details.

\begin{figure}[]
\begin{center}
 \includegraphics[width=8.2cm,angle=0,clip]{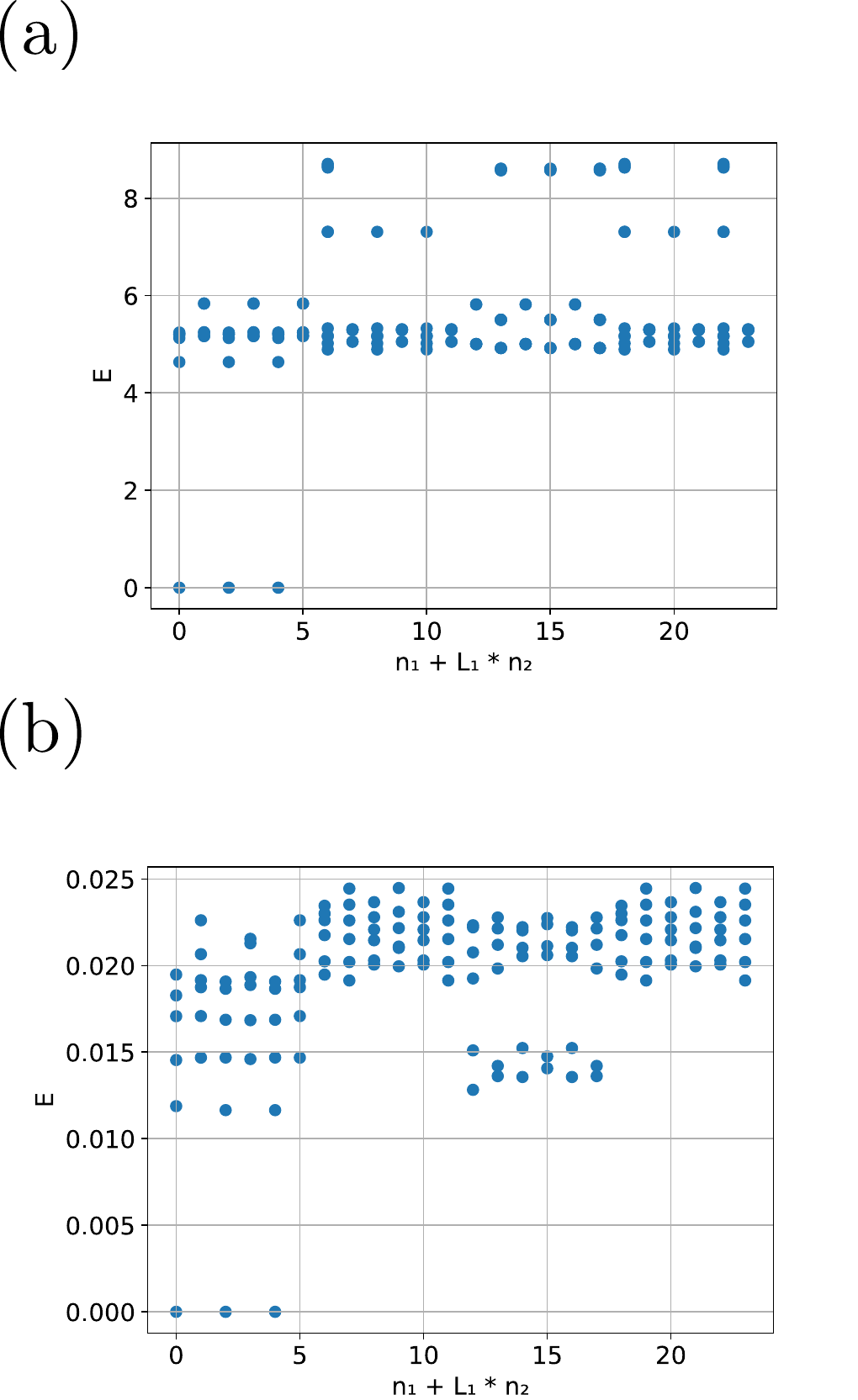}
 \caption{Exact Diagonalizations of the coherent-state model with $U(\bm{\delta})=1$ and $U(\bm{\delta})=\delta_{\bm{\delta},\bm{0}}$. Total momentum sectors are labeled by $(n_1,n_2)$.}
 \label{fig4}
\end{center}
\end{figure}

From the above discussion, it follows that the Hamiltonian (\ref{continuous}) has three zero-energy gapped ground states.
Since $n_{z,0}n_{z,1}$ is positive semi-definite, the existence of zero-energy ground states for the Hamiltonian (\ref{continuous}) implies that they are also zero modes of $n_{z,0}n_{z,1}$.
Using this frustration-free property, one can conclude that the model (\ref{coherent-hubbard}) for arbitrary $U(\bm{\delta})$ has the same zero-energy ground states.
Note that this discussion does not guarantee the absence of other ground states or gapless excited states.
For a natural choice of $U(\bm{\delta})$, however, this model is expected to host three gapped topological ground states.
In Fig. \ref{fig4} (a), we show the exact diagonalization result for $U(\bm{\delta})=1$ with $N_x\times N_y=6\times4$.
Clearly, the typical gapped topological ground states appear at total momenta $\bm{K}=(0,0),(2\times2\pi/6,0),(4\times2\pi/6,0)$.
This gapped nature is also observed in an extreme case, $U(\bm{\delta})=\delta_{\bm{\delta},\bm{0}}$ [Fig.\ref{fig4}(b)].

Note that if the coefficient $a^{(\rm LLL)}_{\bm{k}}(\bm{\delta},i)$ for a given Chern insulator is used instead of $a_{\bm{k}}(\bm{\delta},i)$, one can define a model with exact zero-energy ground states.
However, such basis states are less localized compared to coherent-like states, and the corresponding interaction Hamiltonian would thus be unnatural.

\subsection{Exact diagonalization for Chern insulators\label{edchern}}
\begin{figure*}[]
\begin{center}
 \includegraphics[width=14cm,angle=0,clip]{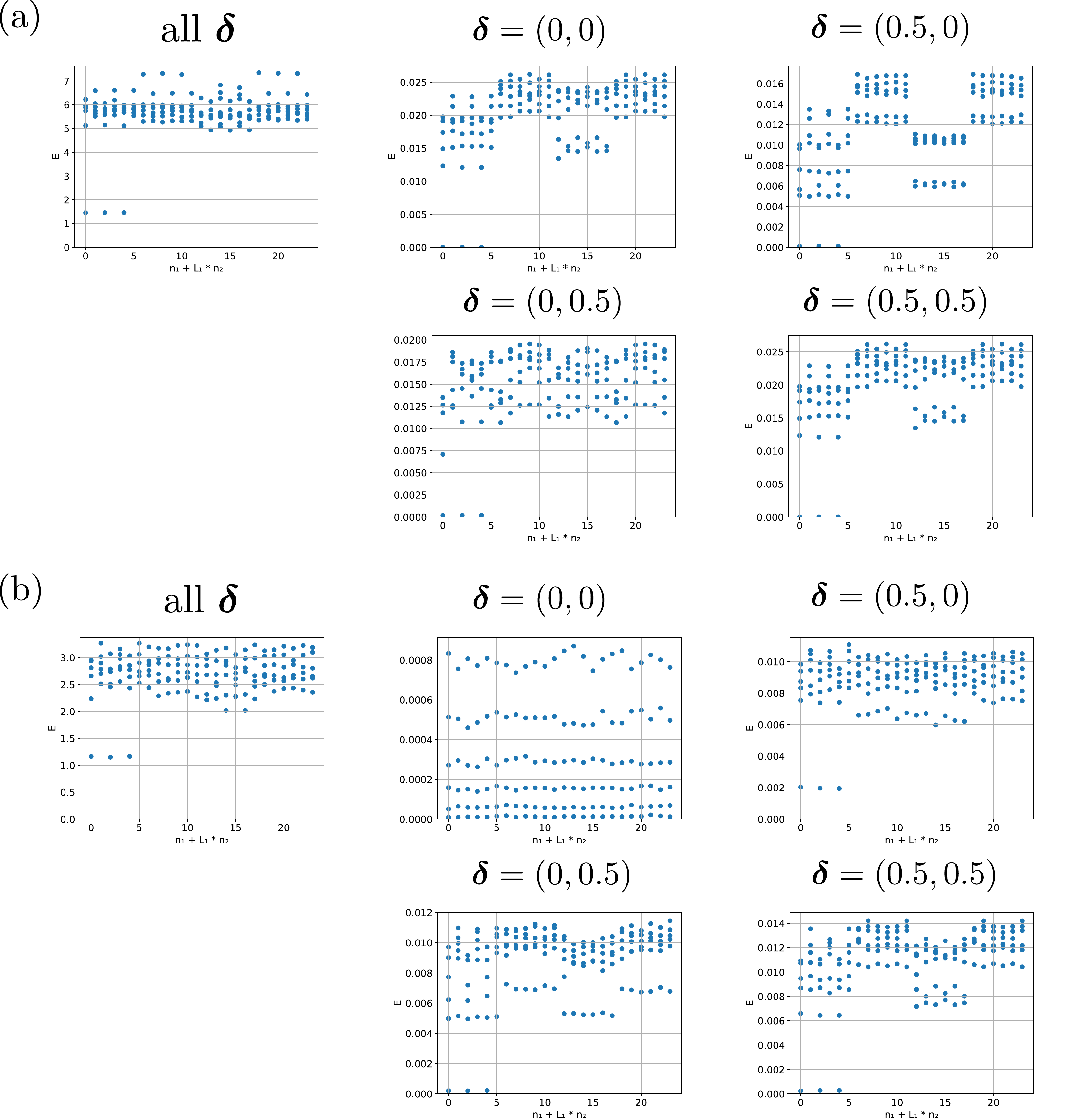}
 \caption{Exact Diagonalizations of the coherent-like-state Hamiltonians for (a) the checkerboard lattice model and (b) the QWZ model. $\bm{\delta}=\bm{0}$ corresponds to  $\tilde{\bm{r}}_1=(0,-0.5),~\tilde{\bm{r}}_2=(-0.5,0)$ for the checkerboard-lattice model and $\tilde{\bm{r}}_1=(0,0),~\tilde{\bm{r}}_2=(0,0)$ for the QWZ model, respectively. }
 \label{fig5}
\end{center}
\end{figure*}

In the case of the coherent-like states, the above connection to the solvable model is absent.
In general, the ground states of the model (\ref{coherent-hubbard}) may have finite energies.
We again consider both the checkerboard-lattice model and the QWZ model.
We perform exact diagonalization for the systems with $6\times4$ unit cells.
For numerical stability, we use the values of $a_{\bm{k}}(\bm{\delta},0)$ obtained from a system with $12\times12$ unit cells, as described in Sec. \ref{overbasis}.
In Fig. \ref{fig5}(a), we show the results for the checkerboard lattice model.
For $U(\bm{\delta})=1$, nearly threefold-degenerate gapped ground states emerge, and their energies are less than one third of that of the first excited state.
In a realistic situation, however, $U(\bm{\delta})$ can depend on the position $\bm{\delta}$.
As an extreme case, we consider a situation where interactions occur only at a specific point within the unit cell.
In this model, the position dependence of the energy spectrum is not significant.
In Fig. \ref{fig5}(b), we show the results for the QWZ model.
For $U(\bm{\delta})=1$, nearly threefold-degenerate gapped ground states appear.
However, the ground state energies are more than half of that of the first excited state.
Moreover, the position dependence of the energy spectrum is nontrivial.
For $U(\bm{\delta})=\delta_{\bm{\delta},\bm{0}}$, the threefold degeneracy collapses into a massively degenerate structure.
These results suggest that the stability of the FCI depends on the interaction details.

Under what conditions do FCI ground states remain insensitive to the microscopic details of the interactions?
In the case of the lowest Landau level, the scattering element exhibits the following symmetry:
\begin{align}
    &V^{\bm{\delta}}_{\bm{k},\bm{q}-\bm{k};\bm{q}-\bm{k}',\bm{k}'}=V^{\bm{0}}_{[\bm{k}],[\bm{q}-\bm{k}];[\bm{q}-\bm{k}'],[\bm{k}']},\notag\\
    &V^{\bm{\delta}}_{\bm{k},\bm{q}-\bm{k};\bm{q}-\bm{k}',\bm{k}'}:=a_{\bm{k}}(\bm{\delta},0)a_{\bm{q}-\bm{k}}(\bm{\delta},1)a^*_{\bm{q}-\bm{k}'}(\bm{\delta},1)a^*_{\bm{k}'}(\bm{\delta},0),
\end{align}
where  $[\bm{p}]=\bm{p}+2\pi\bm{\delta}\times \hat{e}_3$.
This symmetry arises from the fact that the quantum Hall system is defined in a continuum with continuous translational symmetry. Although the Chern insulator system is a lattice system, by introducing $\bm{\delta}$ in this study, it can be discussed within the framework of a continuum system. In other words, a system that mimics this translation invariance of the scattering elements in momentum space is expected to host ground states that are insensitive to the details of the interaction parameters, $U(\bm{\delta})$.

\section{$\mathbb{Z}_2$ coherent-like states\label{z2section}}
A description using coherent-like states is also useful for other topological materials. Here, as an example, we consider a two-dimensional $\mathbb{Z}_2$ topological insulator.
In $\mathbb{Z}_2$ topological insulators, although it is possible to construct exponentially localized Wannier functions, doing so comes at the cost of breaking the Kramers-pair structure.
We will see below that, in $\mathbb{Z}_2$ coherent-like states, this Kramers-pair structure can be preserved.

As we mentioned in the Discussion of Ref. \cite{okuma2024constructing}, $PZ^*P$ possesses an odd number of zero modes if the $\mathbb{Z}_2$ number is nontrivial.
This is because one can consider the $\mathbb{Z}_2$-type index theorem \cite{fukui2009z2} for the corresponding Dirac operator in momentum space, Eq.(\ref{dirac}) \cite{okuma2024constructing}, instead of the conventional index theorem.
In the same way as the construction described above, one can define the coherent-like states for $PZ^*P$.
We denote these states by $\ket{\zeta^{+}_{\bm{R}}}$:
\begin{align}
    PZ^*P\ket{\zeta^+_{\bm{R}}}=Z^*_{\bm{R}}\ket{\zeta^+_{\bm{R}}}.
\end{align}
Moreover, the Kramers counterpart, $\ket{\zeta^{-}_{\bm{R}}}:=T\ket{\zeta^{+}_{\bm{R}}}^*$, is an eigenstate of $PZP$:
\begin{align}
    PZP\ket{\zeta^-_{\bm{R}}}=Z_{\bm{R}}\ket{\zeta^-_{\bm{R}}}.
\end{align}
For a general non-Hermitian matrix $N$ with the fermionic transpose-type time-reversal symmetry \cite{kawabata2019symmetry}, $TN^TT^{-1}=N$ $(TT^*=-1)$, the following holds:
\begin{align}
    &N\ket{\lambda}=\lambda\ket{\lambda}\notag
    \\
    \Rightarrow& N^{\dagger}(T\ket{\lambda}^*)=(T^{-1}NT)^{*}T\ket{\lambda}^*=-(T^{-1})^*N^*\ket{\lambda}^*\notag\\
    &=T(N\ket{\lambda})^*=\lambda^{*}(T\ket{\lambda}^*).
\end{align}
In the present case, $PZ^*P$ obeys the same transverse-type time-reversal symmetry if the Hamiltonian, or the projection operator, obeys the fermionic time-reversal symmetry, $TH^*T^{-1}=H$ $(TP^*T^{-1}=P)$ with $TT^*=-1$, and the matrix $T$ satisfies $[T,Z^*]=0$.
In a natural setup, $T$ describes the spin degrees of freedom and commutes with the vortex function that consists of the position operators.

In summary, in the case of the $\mathbb{Z}_2$ topological insulator, coherent-like states are defined as Kramers pairs, which are orthogonal to each other and localized at the same position.
When spin is conserved, these states can be understood as coherent-like states of Chern bands with Chern number $\pm1$. In such cases, coherent-like states belonging to different spin sectors on different sites are orthogonal to each other.
By contrast, in a general $\mathbb{Z}_2$ topological insulator, spin is not conserved, and this orthogonality is lost. As we have already seen, the interplay between non-orthogonality and interactions is closely connected to exotic physical properties. Therefore, the difference in strongly correlated phenomena between spin-conserving and spin-nonconserving systems is an interesting topic.

\section{Discussion\label{discussionsection}}
In this section, we discuss several remaining issues.
In this study, we considered only the interaction Hamiltonian projected onto the Chern band. On the other hand, effects originating from the single-particle Hamiltonian are also expected to be important in real materials.
As the lowest-order correction, one possible approach is to project the single-particle states onto the Chern band while retaining only the shape of the dispersion in the energy.
In the wavepacket of the coherent-like states, depending on the position $\bm{\delta}$, the weight is concentrated near the corresponding momentum.
This implies that when dispersion is present, the coherent-like states experience a single-particle energy that depends on the localization position.
In other words, dispersion breaks translational symmetry both in momentum space and in real space.
As discussed in Sec. \ref{edchern}, the continuous symmetry with respect to $\bm{\delta}$ plays a crucial role in stabilizing the FCI.
Therefore, the breaking of spatial translational symmetry due to dispersion implies an instability of the FCI.
Investigating dispersion effects in real materials based on this picture is an important direction for future studies.

Beyond Chern insulators, the investigation of strongly correlated topological materials using coherent-like states remains an important open issue. As already mentioned, in $\mathbb{Z}_2$ topological insulators, coherent-like states are introduced in the form of Kramers doublets. Since two localized states are defined at the same spatial position, this setting is expected to give rise to richer and more complex quantum phases. Understanding such phases may benefit from the use of a localized basis.

\acknowledgements
I gratefully acknowledge valuable discussions with Tomonari Mizoguchi and Ikuma Tateishi.
This work was supported by JSPS KAKENHI Grant No.~JP20K14373 and No.~JP23K03243.
\\
\appendix
\section{Momentum representation of coherent states in the lowest Landau level\label{xlqmomentum}}
In Ref. \cite{qi2011generic}, the coherent state on cylinder is defined as
\begin{align}
    |z=x+iy\rangle\propto\sum_{n\in\mathbb{Z}}e^{-i\frac{2\pi n}{N_y}y}e^{-\pi\left(x-\frac{n}{N_y}\right)^2}\left|p_y=\frac{2\pi n}{N_y}\right\rangle,\label{xlqcoherent}
\end{align}
where $p_y$ is the momentum in the $y$ direction.
In addition to this definition, slightly different definitions are also used in the literature, for example, in Ref. \cite{ishikawa1999field}.
For simplicity, we first consider $x=y=0$.
The momentum can be divided into the crystal momentum, $k_y$, and the unit-cell index, $R_x$:
\begin{align}
    p_y=2\pi R_x+k_y.
\end{align}
Using this decomposition, the momentum basis can be regarded as a Wannier basis localized only in the x-direction:
\begin{align}
    \left|p_y=\frac{2\pi n}{N_y}\right\rangle=\ket{k_y,R_x}.
\end{align}

Then, we obtain
\begin{align}
    |0\rangle&\propto\sum_{k_y,R_x}e^{-\pi\left(R_x+\frac{ky}{2\pi}\right)^2}\left|k_y,R_x\right\rangle\notag\\
    &\propto\sum_{\bm{k}}\sum_{R_x}e^{-\pi\left(R_x+\frac{ky}{2\pi}\right)^2}e^{-ik_xR_x}\ket{\bm{k}}\notag\\
    &=\sum_{\bm{k}}e^{-\frac{k_y^2}{4\pi}}\sum_{R_x}e^{-\pi R_x^2}\exp\left[2\pi iR_x\left(\frac{-k_x+i~k_y}{2\pi}\right)\right]\ket{\bm{k}}\notag\\
    &=:\sum_{\bm{k}}a^{(\rm LLL)}_0(\bm{k})\ket{\bm{k}},
\end{align}
where
\begin{align}
    a^{(\rm LLL)}_0(\bm{k})=e^{-\frac{k_y^2}{4\pi}}\vartheta_3\left(\frac{-k_x+i k_y}{2\pi}|i\right).
\end{align}
Here, $\vartheta_3(z|\tau)$ is the elliptic theta function.
The coefficient $a_{0}^{(\rm{LLL})}$ is the zero mode of $ z_{\bm{k}}^{\dagger}$ with $\bm{A}_{\bm{k}}=(-k_y/2\pi,0)$, which corresponds to the constant Berry curvature in momentum space:
\begin{align}
    [\hat{z}^{(\mathrm{LLL})}_{\bm{k}}]^{\dagger}~a_{0}^{(\rm{LLL})}(\bm{k})=0.
\end{align}

Similarly, we can write down the momentum-space representation for general $(x,y)$: 
\begin{align}
        |z\rangle\propto\sum_{\bm{k}}\sum_{R_x}e^{-i(k_y+2\pi R_x)y}e^{-\pi\left(x-\frac{ky}{2\pi}-R_x\right)^2}e^{-ik_xR_x}\ket{\bm{k}}.\label{explicit}
\end{align}
By performing the same calculation, we obtain the following expression for the case $(x,y)=(\delta_x,\delta_y)$:
\begin{align}
    &a^{(\rm LLL)}_{\bm{k}}(\bm{\delta},0)=e^{-ik_y\delta_y}a^{(\rm LLL)}_0(\bm{k}+2\pi\bm{\delta}\times \hat{e}_3),\\
    &
     [\hat{z}^{(\mathrm{LLL})}_{\bm{k}}]^{\dagger}a^{(\rm LLL)}_{\bm{k}}(\bm{\delta},0)=(\delta_x-i\delta_y)a^{(\rm LLL)}_{\bm{k}}(\bm{\delta},0).
\end{align}
Clearly, the last expression is compatible with our definition of the coherent state:
\begin{align}
    P_{\rm LLL}z^*P_{\rm LLL}\ket{\delta_x+i\delta_y}=(\delta_x+i\delta_y)^*\ket{\delta_x+i\delta_y}.
\end{align}

Note that these expressions can be used for the torus geometry by enforcing $\ket{k_y,R_x}=\ket{k_y,R_x+N_x}$ and $k_x=2\pi n/N_x$.

\section{Proof of the equivalence between the Hamiltonians (\ref{continuous}) and (\ref{localham})\label{proof1}}
We here prove the equivalence between the Hamiltonians (\ref{continuous}) and (\ref{localham}).
We first consider the momentum-space representation.
By using Eq. (\ref{explicit}), we get the following relation:
\begin{align}
    &\partial_{\delta_x}a^{(\rm LLL)}_{\bm{k}}(\bm{\delta},0)=-2\pi(\delta_x-\frac{k_y}{2\pi}-R_x)a^{(\rm LLL)}_{\bm{k}}(\bm{\delta},0)\notag\\
    &=\pi\left[\hat{z}_{\bm{k}}^{\rm(LLL)}-(\delta_x+i\delta_y)\right]a^{(\rm LLL)}_{\bm{k}}(\bm{\delta},0)=\sqrt{\pi}a^{(\rm LLL)}_{\bm{k}}(\bm{\delta},1).
\end{align}
We recall the definition of the field operator:
\begin{align}
    c^{\dagger}_{\bm{R},\bm{\delta},m}=\sum_{\bm{k}}\frac{e^{-i\bm{k}\cdot\bm{R}}}{\sqrt{N_{\rm unit}}}a_{\bm{k}}(\bm{\delta},m)c^{\dagger}_{\bm{k}}.
\end{align}
By replacing $\partial_{\delta_x}$ with $\partial_x$, we obtain
\begin{align}
    \partial_{x}c^{\dagger}_{z,0}=\sqrt{\pi}~c^{\dagger}_{z,1},~\partial_{x}c_{z,0}=\sqrt{\pi}~c_{z,1}.
\end{align}
Similarly, we obtain
\begin{align}
    \partial_{y}c^{\dagger}_{z,0}=-i\sqrt{\pi}~c^{\dagger}_{z,1}-2\pi i\delta_xc^{\dagger}_{z,0},\notag\\
    \partial_{y}c_{z,0}=+i\sqrt{\pi}~c_{z,1}+2\pi i\delta_xc_{z,0}.
\end{align}
Thus, we get 
\begin{align}
    &c_{z,0}(-i\partial_x)c_{z,0}=-i\sqrt{\pi}c_{z,0}c_{z,1},\\
    &c_{z,0}(-i\partial_y)c_{z,0}=\sqrt{\pi}c_{z,0}c_{z,1}.
\end{align}
In the second line, we have used $c^2_{z,0}=0$.
Clearly, these equations indicate the equivalence between (\ref{continuous}) and (\ref{localham}), up to an overall constant.

\section{Proof of the existence of exact zero modes in the model on the lowest Landau level\label{proof}}
Reference \cite{qi2011generic} considered the following model on a cylinder:
\begin{align}
    H\propto\int dxdy ~\bm{b}^{\dagger}(z)\cdot\bm{b}(z),~\bm{b}(z)=c(z)(-i\nabla)c(z),\label{xlqmodel}
\end{align}
where $c(z)$ is the annihilation operator of the coherent state (\ref{xlqcoherent}).
They showed that this model can be rewritten as the following second-quantized form of Haldane’s pseudopotential for filling factor $\nu=1/3$:
\begin{align}
    H\propto\sum_{j,k,l}\eta_k~\eta_l~c^{\dagger}_{j+k}c^{\dagger}_{j-k}c_{j-l}c_{j+l},\label{dung-hai}
\end{align}
where $\eta_l=l~e^{-2\pi l^2/N_y^2}$.
Since $j\pm k,~j\pm l$ are integers, the summation runs over the triples $(j,k,l)$ in which all entries are integers or all entries are half-odd integers. 
The model (\ref{dung-hai}) corresponds to the case $\kappa=2\pi l_B/N_y$ and $2\pi l^2_B=1$ in Ref. \cite{lee2004mott}, and it possesses a unique zero-energy gapped ground state.

In the following, we consider the model (\ref{xlqmodel}) on the torus geometry and prove that it can be rewritten as a model in Ref. \cite{ortiz2013repulsive}, which has threefold-degenerate zero-energy ground states.
We first impose the periodic boundary condition $c_j=c_{j+N}$.
Since the proof of the correspondence between (\ref{xlqmodel}) and (\ref{dung-hai}) has nothing to do with the condition about $c_j$, it also holds for the torus geometry.
Due to the periodicity, the following holds for an arbitrary $s\in \mathbb{Z}$:
\begin{align}
    &\sum_{j,k,l}\eta_k~\eta_l~c^{\dagger}_{j+k-sN}c^{\dagger}_{j-k-sN}c_{j-l}c_{j+l}\notag\\&=\sum_{j,k,l}\eta_{k+sN}~\eta_l~c^{\dagger}_{j+k}c^{\dagger}_{j-k}c_{j-l}c_{j+l}.
\end{align}
A similar relation holds for the annihilation operators.
By summing over such equivalent models and taking the average,
we get
\begin{align}
    &H\propto\sum_{j,k,l}\eta^{(\rm torus)}_k~\eta^{(\rm torus)}_l~c^{\dagger}_{j+k}c^{\dagger}_{j-k}c_{j-l}c_{j+l},\\
    &\eta^{(\rm torus)}_k=\sum_{s\in\mathbb{Z}}\eta_{k+sN}.
\end{align}
The new coefficient $\eta^{(\rm torus)}_k$ is also antisymmetric:
\begin{align}
    \eta^{(\rm torus)}_{-k}&=\sum_{s\in\mathbb{Z}}\eta_{-k+sN}=-\sum_{s\in\mathbb{Z}}\eta_{k-sN}\notag\\
    &=-\sum_{s\in\mathbb{Z}}\eta_{k+sN}=\eta^{(\rm torus)}_{k}.
\end{align}

Since this expression is redundant, we will henceforth consider restricting the range of the summation over triples $(j,k,l)$.
The $k=0$ terms can trivially be omitted.
For $k<0$ terms, the following holds:
\begin{align}
    &\eta^{(\rm torus)}_kc^{\dagger}_{j+k}c^{\dagger}_{j-k}=\eta^{(\rm torus)}_{-|k|}c^{\dagger}_{j-|k|}c^{\dagger}_{j+|k|}\notag\\
    &=-\eta^{(\rm torus)}_{-|k|}c^{\dagger}_{j+|k|}c^{\dagger}_{j-|k|}=\eta^{(\rm torus)}_{|k|}c^{\dagger}_{j+|k|}c^{\dagger}_{j-|k|}.
\end{align}
A similar discussion holds for $l$.
Thus, we can set $0<k,l$.
The $k=N/2$ terms can be omitted because $c^{\dagger}_{k+N/2}c^{\dagger}_{k-N/2}=(c^{\dagger}_{k+N/2})^2=0$.
For $N/2<k<N$, the following holds:
\begin{align}
    &\eta^{(\rm torus)}_kc^{\dagger}_{j+k}c^{\dagger}_{j-k}=\eta^{(\rm torus)}_kc^{\dagger}_{j-(N-k)}c^{\dagger}_{j+(N-k)}\notag\\
    &=\eta^{(\rm torus)}_{-k}c^{\dagger}_{j+(N-k)}c^{\dagger}_{j-(N-k)}=\eta^{(\rm torus)}_{N-k}c^{\dagger}_{j+(N-k)}c^{\dagger}_{j-(N-k)}.
\end{align}
Thus, these terms are equivalent to $0<k':=N-k<N/2$.
The $k=N$ terms can be omitted because $c^{\dagger}_{k+N}c^{\dagger}_{k-N}=(c^{\dagger}_{k})^2=0$.
Because of the periodicity, the $k>N$ terms are equivalent to the terms discussed above.
A similar discussion holds for $l$.
Thus, we can set $0<k,l<N/2$.
Finally, we consider the range of $j$.
Because of the periodicity, we can set $1/2\leq j\leq N-1/2$, or equivalently, $0<j<N$.
Thus, we obtain the following model:
\begin{align}
    H\propto\sum_{0<j<N}\sum_{0<j,l<N/2}\eta^{(\rm torus)}_k~\eta^{(\rm torus)}_l~c^{\dagger}_{j+k}c^{\dagger}_{j-k}c_{j-l}c_{j+l}.
\end{align}
By changing $l_B=1/\sqrt{2\pi}$ to $l_B=1$, we get the model in Ref. \cite{ortiz2013repulsive}.

\section{Models\label{models}}
In this section, we write down the Bloch Hamiltonians for the Chern insulators used in the main text.
\subsection{Checkerboard-lattice model}
This model was studied in terms of the fermionic FCI \cite{Neupert-Santos-Chamon-Mudry-11}. 
The Bloch Hamiltonian matrix is given by
\begin{align}
    H(\bm{k})=
    \begin{pmatrix}
    2t_2(\cos k_1-\cos k_2)&t_1f^*(\bm{k})\\
    t_1f(\bm{k})&-2t_2(\cos k_1-\cos k_2)\\
    \end{pmatrix},
\end{align}
where
\begin{align}
    &f(\bm{k})=e^{-i\pi/4}\left[1+e^{i(k_2-k_1)}\right]+e^{i\pi/4}\left[e^{-ik_1}+e^{ik_2}\right].
\end{align}
In numerical calculations, we set $t_1=1,t_2=\sqrt{2}/2$. For this parameter, the band structure becomes almost flat.

\subsection{Qi-Wu-Zhang (QWZ) model}
This model was proposed in Ref.~\cite{Qi-Wu-Zhang-06} and is also referred to as the Wilson-Dirac model. 
The Bloch Hamiltonian matrix is given by
\begin{align}
    &H(\bm{k})=\sin k_1 \sigma_x+\sin k_2 \sigma_y+(m-\cos k_1-\cos k_2)\sigma_z, \label{eq:QWZ_Bloch}
\end{align}
where $\sigma_{x,y,z}$ are the Pauli matrices. 
In numerical calculations, we set $m=1$.

\bibliography{FCI}
\end{document}